\documentstyle[aps,twocolumn,epsfig]{revtex}
\newcommand{\beq}{\begin{equation}}
\newcommand{\eeq}{\end{equation}}
\newcommand{\bea}{\begin{eqnarray}}
\newcommand{\eea}{\end{eqnarray}}
\newcommand{\ba}{\begin{array}}
\newcommand{\ea}{\end{array}}
\newcommand{\bc}{\begin{center}}
\newcommand{\ec}{\end{center}}

\newcommand{\etal}{{\it et al.}}
\newcommand{\bml}{\begin{mathletters}}
\newcommand{\eml}{\end{mathletters}}
\newcommand{\commentout}[1]{{}}
\newcommand{\k}{{\bf k}}

\newcommand{\I}{{\cal I}}
\newcommand{\q}{{\bf q}}

\newcommand{\half}{\hbox{$1\over2$}}

\newcommand{\eq}[1]{(\ref{#1})}
\newcommand{\tr}{\text{Tr}}
\newcommand{\cpv}{\text{P}}
\newcommand{\adag}{a^\dagger}
\newcommand{\bdag}{b^\dagger}
\newcommand{\vol}[1]{{\bf #1}}

\begin{document}
\draft
\flushbottom
\wideabs
{
\title{Anomalous Frequency Shift in Photoassociation of a
Bose-Einstein Condensate}
\author{Matt Mackie}
\address{Helsinki Institute of Physics, PL 64, FIN-00014
Helsingin yliopisto, Finland}

\date{\today}

\maketitle

\begin{abstract}
We theoretically investigate the effect of anomalous quantum
correlations on the light-induced frequency shift in the photoassociation spectrum
of a Bose-Einstein condensate (BEC). Anomalous quantum correlations arise
because, although formed from a pair of zero-momentum condensate atoms, a
condensate molecule need not dissociate back to the BEC, but may just as well form
a noncondensate atom pair with equal and opposite momentum, i.e., due to rogue
photodissociation. The uncorrelated frequency shift of the photoassociation
spectrum is to the red and linearly dependent on the laser intensity $I$. In
contrast, anomalous correlations due to rogue dissociation lead to a
blueshifted photoassociation spectrum. For sufficiently low light intensities, the
rogue blueshift is dominant and proportional to $\sqrt{I}$. 
\end{abstract}
}

Photoassociation occurs when two free atoms absorb a laser photon, thereby
jumping from the two-atom continuum to a bound molecular state\cite{WEI99}. If the
atoms form a Bose-Einstein condensate (BEC), then the molecules will
too\cite{JAV99,HEI00,HOP01,VAR01}. While evidence for the photoassociative
formation of quantum degenerate molecules has yet to emerge, the quest has led to
the observation\cite{WYN00} of strongly enhanced molecule
formation\cite{BUR97,JAV98,JUL98}, precise measurements\cite{GER01,MCK02,PRO03} of
the light-induced shift of the photoassociation spectrum\cite{JAV98,FED96,BOH99},
and tests\cite{MCK02,PRO03} of a fundamental (non-unitary) limit to the
atom-molecule conversion rate\cite{JAV99,KOS00,JAV02}.

In coherent photoassociation, the initial
atoms belong to a Bose-Einstein condensate, and therefore so will the
molecules. However, transitions to the continuum of noncondensate atomic
modes can occur because photodissociation of a zero-momentum ($\k=0$) condensate
molecule need not take the atoms back to the $\k=0$ atomic condensate,
but may just as well end up with two atoms with opposite momenta
($\pm\k$), i.e., due to rogue~\cite{JAV99,KOS00,JAV02} or
unwanted~\cite{GOR01,HOL01} photodissociation. Rogue dissociation to
noncondensate modes ultimately leads to anomalous quantum
correlations, which are the bosonic equivalent of the Cooper pairs responsible for
superconductivity. An immediate consequence of said pairing is the above-mentioned
non-unitary limit on the rate of conversion from an atomic to a molecular
condensate\cite{JAV99,KOS00,JAV02}; additionally, there is the
possibility of creating strongly-correlated twin atomic beams\cite{KHE02}.  

The purpose of this Letter is to reveal the effect of anomalous
quantum correlations on the light-induced shift of the photoassociation spectrum of
a Bose-Einstein condensate. Uncorrelated free-bound couplings
necessarily introduce a redshift to the photoassociation spectrum which, to lowest
nontrivial order, is linear in laser
intensity~\cite{JAV98,GER01,MCK02,PRO03,FED96,BOH99}. In contrast, we find that
anomalous correlations lead to a blueshifted photoassociation spectrum. To lowest
nontrivial order, the rogue frequency shift is dominant and
proportional to the squareroot of the photoassociation laser intensity.

Our description starts with a plane-wave laser field with photon momentum $\q$ to
drive photoassociation and photodissociation. Initially there is only a condensate
of $N$ zero-momentum atoms present, and these are characterized by the boson
operator $a\equiv a_0$. By momentum conservation, only molecules with momentum $\q$
will be generated in the primary photoassociation process, and these are
characterized by the boson operator $b\equiv b_\q$. Upon photodissociation, the
molecules break up into pairs of atoms with equal and opposite momenta, and these
atoms are characterized by the boson operators $a_{\pm\k}$. The Hamiltonian
for this system is given by
\bea
\hbar^{-1}H &=& \delta\bdag b + \half\, \sum_\k \epsilon_k \adag_\k a_\k
  +{\Omega\over\sqrt{N}}(\bdag aa+\adag\adag b), \nonumber\\
&&+{\Omega\over\sqrt{N}}\sum_{\k\neq0}f(\epsilon_k)\left[
  \bdag a_\k a_{-\k}+\adag_{-\k}\adag_\k b\,\right].
\eea

Here the quantity
$\delta$ is the detuning of the laser above the photodissociation
threshold of the molecules, which is corrected for photon recoil
effects\cite{JAV99,KOS00}, and which is a measure of the molecular binding energy.
The atom-molecule Rabi coupling $\Omega$ is given by the expression
\beq
\Omega=\lim_{\epsilon_k\rightarrow0}
  \sqrt{{\sqrt{2}\,\pi\hbar^{3/2}n\over\mu^{3/2}}\,
    {\Gamma_0(\epsilon_k)\over\sqrt{\epsilon_k}}},
\label{BEC_RABI}
\eeq
where $n=n_A+2n_M$ is an invariant condensate density, derived from the densities of
atoms and molecules,
$\mu=m/2$ is the reduced mass of an atom pair, and $\Gamma_0$ is the rate of
photodissociation to an atom pair of energy
$\hbar\epsilon_k=\hbar^2k^2/2\mu$. In order for Eq.~\eq{BEC_RABI}
to give a finite atom-molecule coupling, for low energies
$\Gamma_0\propto\sqrt{\epsilon_k}$ must hold true, which is
just another way of stating the Wigner threshold law\cite{WEI99}. Lastly, the
coupling of the molecular condensate mode to noncondensate atomic modes is
$\Omega f(\epsilon_k)/\sqrt{N}$, which is a measure of the coupling strength for
photodissociation of a molecule to an atom pair of energy $\hbar\epsilon_k$,
with $f(\epsilon_k)$ describing the energy (wavenumber) dependence of the
coupling.

Next we describe rogue dissociation in the language of system-reservoir
interactions~\cite{TEXT}. The intracondensate coupling term is neglected, since
it is not necessary to the main point, and since it can be re-introduced
later anyway\cite{TEXT}. The focus is now on a molecular condensate coupled to a
bath of dissociated atom pairs. Switching to the interaction picture, the
system-reservoir Hamiltonian is
\beq
\hbar^{-1}H_I(t)=
  \bdag\Gamma(t)e^{i\delta t}+\Gamma^\dagger(t)be^{-i\delta t}\,
\eeq
with the reservoir operator $\Gamma(t)$ defined as
\beq
\Gamma(t) = {\Omega\over\sqrt{N}}\sum_{\k\neq0}
  f(\epsilon_k)\,a_\k a_{-\k} e^{-i\epsilon_kt}.
\label{ROGUE_OP}
\eeq
The equation of motion for the total density matrix is then
$\dot\rho_T=-i\hbar^{-1}[H_I(t),\rho_T(t)]$.
The reduced density matrix for the system is obtained by
integrating out (tracing over) the reservoir degrees of freedom:
$\rho(t)=\tr\left\{\rho_T(t)\right\}_R=\left\langle\rho_T(t)\right\rangle_R$.
It is assumed that, initially, there are no correlations between the
system and reservoir:
$\rho(0)=\left<\rho_T(0)\right>_R=\left<\rho(0)\otimes\rho_R\right>_R$. The
so-called master equation is then the equation of motion for the reduced density
matrix. To second order in perturbation theory, the master equation for rogue
dissociation is
\beq
\dot\rho(t)=U(t)\rho(t),
\label{ROGUE_MASTER}
\eeq
with the generator of time evolution given as
\bea
U(t)&=&
  -{i\over\hbar}\left<[H_I(t),\rho_R\otimes(\cdot)]\right>_R
\nonumber\\
    &&+{1\over\hbar^2}\int_0^t dt'\, \left<[H_I(t'),\rho_R\otimes
      \left<[H_I(t'),\rho_R\otimes(\cdot)]\right>_R]\right>_R
\nonumber\\
    &&-{1\over\hbar^2}\int_0^t dt'\,
      \left<\left[H_I(t),[H_I(t'),\rho_R\otimes(\cdot)]\right]\right>_R.
\label{DEFU}
\eea

The departure from a textbook
treatment~\cite{TEXT} lies in the rogue dissociation paradigm
$\left<\Gamma(t)\right>\neq0$, meaning that
$\left<[H_I(t),\rho_R\otimes\rho]\right>_R$ is nonvanishing and first order terms
are {\em not absent} from the master equation~\eq{ROGUE_MASTER} for the reduced
density matrix.
For small couplings, only terms
$\propto\left<\Gamma(t)\right>_R$ will contribute, and the generator of time
evolution simplifies to
\bml
\bea
U(t)=\I [\bdag,(\cdot)] - \I^*[b,(\cdot)], \\
  \I = -i{d\over dt}\int_0^t dt'\,\left<\Gamma(t')\right>_R
    e^{i\delta t'}.
\eea
\eml
The quantum-correlated shift of the molecular binding
energy is buried in the integral $\I$, and we now proceed with its extraction.
Substitution of Eq.~\eq{ROGUE_OP} yields
\beq
\I=-i{d\over dt}\int_0^t dt'\,
  {\Omega\over\sqrt{N}}\sum_{\k\neq0}f(k)
    \left<a_\k a_{-\k}\right>_R e^{i(\epsilon_k-\delta)t'}.
\eeq
The summation of momentum states is converted to an integral over frequency
according to\cite{NO_ZERO}
\beq
{1\over N}\sum_{\k\neq0}G(k)\rightarrow{1\over4\pi^2\omega_\rho^{3/2}}\,
  \int_0^\infty d\epsilon \sqrt{\epsilon}\,G(\epsilon),
\eeq
where $\omega_r=\hbar n^{2/3}/2\mu$ is our usual density-dependent
frequency parameter\cite{JAV99,KOS00,JAV02}. Next we recall that
$f(\epsilon)$ contains the wavenumber dependence of the system-reservoir
coupling, and choose a Wigner threshold coupling, $f(\epsilon)=1$, for
all energies. Also, as was implicit to Ref.\cite{JAV02}, the rogue pair
correlation function is written
$C(\epsilon)=\left<a_\k a_{-\k}\right>
  \sqrt[4]{\epsilon}/\left(2\sqrt{\pi}\omega_r^{3/4}\right)$, which has
units of $(\rm{frequency})^{-1/2}$. Lastly, the system-reservoir Rabi coupling is
redefined as
$\xi=\Omega/(2\sqrt{\pi}\omega_r^{3/4})$, which is related to the on-shell
photodissociation rate $\Gamma_0=\xi^2\sqrt{\delta}$.

For $\delta=|\delta|\sim\epsilon_b$,
with the molecular binding energy $\epsilon_b$ taken as the largest frequency in the
problem, the integral $\I$ then becomes
\bml
\beq
\I =
  \half\sqrt{N\Gamma_0}\,\dot\varphi(\epsilon_b)\left|C(\epsilon_b)\right|
    -i\Delta\epsilon_b,
\eeq
\beq
\Delta\epsilon_b =
  {\displaystyle{\sqrt{N}\,\xi\over2\pi}}\,
    \cpv{\displaystyle\int_{-\infty}^\infty} d\epsilon\,
      {\displaystyle{1\over\epsilon}}\,\sqrt[4]{\epsilon+\epsilon_b}\,
        \dot\varphi(\epsilon+\epsilon_b)\left|C(\epsilon+\epsilon_b)\right|,
\label{SHIFT}
\eeq
\eml
where P denotes the Cauchy principle value and we have assumed
$C(\epsilon;t)=|C(\epsilon)|\exp[-i\varphi(\epsilon;t)/2]$ with
$\ddot\varphi\approx 0$. The sought-after shift appears explicity from time
evolution of the mean-field MBEC amplitude
$\dot\beta=\langle b\dot\rho\rangle=-\I^*\beta$, which has a
Schr\"odinger-picture solution
\beq
\beta(t)=\beta(0)e^{-i(\epsilon_b+\Delta\epsilon_b)t}
  e^{-\gamma t/2},
\label{MFA}
\eeq
with $\gamma=\sqrt{N\Gamma_0}\,\dot\varphi(\epsilon_b)|C(\epsilon_b)|$. From the
expression~\eq{MFA}, the frequency shift due to rogue quantum correlations is
clearly to the blue and it is proprtional to
$\sqrt{\Gamma_0}\propto\sqrt{I}$. Note the collective enhancement
factor $\sqrt{N}$, which would play a role if the system were initially a Bose
condensate of molecules, as opposed to a BEC of atoms.

For contrast, we briefly describe the
uncorrelated redshift, borrowing the Ref.~\cite{GER01} account of
the Bohn-Julienne theory~\cite{BOH99}. This shift is a
consequence of the light-induced coupling between the free-atom continuum and the
discrete bound molecular state. Based on Fano theory\cite{FAN61}, valid for any
discrete level coupled to any continuum, the molecular binding energy
is shifted by an amount
\beq
\hbar\Delta\epsilon_b^{(u)} \propto
  \int d\epsilon'\,{D(\epsilon')\over(\epsilon-\epsilon')},
\label{RED}
\eeq
where $D(\epsilon')$ is the density of continuum states at the energy
$\hbar\epsilon'$. The integrand in Eq.~\eq{RED} is of course positive (negative) for
$\epsilon'<\epsilon$ ($\epsilon'>\epsilon$), and the density of states increases
with increasing $\epsilon'$. The negative part of the integrand therefore
contributes more strongly, and the continuum shift $\Delta\epsilon_b$ is always to
the red. This redshift is linearly proportional to the square of the free-bound
coupling, and thus linearly proportional to intensity.

In order for the rogue blueshift to dominate the uncorrelated redshift~\eq{RED},
lowest-order perturbation theory must be valid. According to our previous
work~\cite{KOS00}, the photon recoil frequency,
$\epsilon_R=\hbar/2m\lambdabar^2$, is expected to set the scale for the coupling
strength:
\beq
{\Omega\over\epsilon_R}=\sqrt{n\lambdabar^3}\,\sqrt{I\over I_0}\,,
\label{RABI_SCALE}
\eeq
where
$2\pi\lambdabar$ is the wavelength of the photoassociation laser; also,
$I_0=v\lambda_D^3/(32\pi\lambdabar\tilde{K})$ is the characteristic intensity of
photoassociation to a given molecular level, with $v$ the relative velocity of the
colliding atoms, $\lambda_D$ the deBroglie wavelength (in terms of the reduced
mass), and
$\tilde{K}=dK/dI$ the photoassociation rate coefficient\cite{DK_DI}. The rogue
master equation~\eq{ROGUE_MASTER} is thereby written in dimensionless form, and
first order perturbation theory is expected to be valid for
$\Omega/\epsilon_R\ll 1$, which translates into intensities satisfying
$[(n\lambdabar^3)(I/I_0)]^{1/2}\ll 1$.

In photoassociation of an atomic BEC, i.e., absent a macroscopic number of
molecules, the size of the rogue blueshift should be given roughly by
$\Delta\epsilon_b\sim\gamma/\sqrt{N}\sim\Omega\epsilon_b^{1/2}\epsilon_R/4\pi\omega_r^{3/2}$,
where we have taken $\dot\varphi\sim\epsilon_R$ and
$|C|\sim\sqrt[4]{\epsilon_b}/(2\sqrt{\pi}\omega_r^{3/4})$. Inserting
Eq.~\eq{RABI_SCALE}, the magnitude of the blueshift becomes
\beq
{\Delta\epsilon_b\over\epsilon_R}\sim{1\over4\pi}\,\sqrt{n\lambdabar^3}\,
  \sqrt{ {I\over I_0}\,{\epsilon_b\epsilon_R^2\over\omega_r^3}}\,.
\eeq
To make the density dependence explicit, we write the shift in terms of the
photodissociation rate,
$\Delta\epsilon_b\sim\Gamma_0/[(n\lambdabar^3)(I/I_0)]^{1/2}$.
Besides being proportional to $\sqrt{I}$, the magnitude of the shift is
proportional to $1/\sqrt{n}$, so that smaller condensates will give larger
blueshifts.

Note that second-order perturbation theory is required when
$\Omega/\epsilon_r\sim 1$, which is the regime of experiments performed so
far\cite{GER01,MCK02,PRO03}. In this case, terms such as
$\left<\Gamma\right>^2$ and
$\left<\Gamma\Gamma\right>$ will contribute blueshifts as well. These terms will
add to the usual uncorrelated redshift~\eq{RED},
which arises here from terms like $\left<\Gamma^\dagger\Gamma\right>$.
A quantitative analysis of the higher-order contributions to the frequency shift is
given elsewhere\cite{MAC04}.

In conclusion, low-intensity rogue photodissociation should result in a blueshift
of the molecular binding energy, proportional to $\sqrt{I}$, that dominates the
usual redhshift. The magnitude of this blueshift is proprortional to
$1/\sqrt{n}$, so that smaller condensates will yield more significant shifts.
If the systems starts out as a molecular Bose-Einstein condensate, then the
frequency shift, correlated or not, should reveal the effects of collective
enhancement. Realizing that the anamolous pair correlations herein are the bosonic
equivalent of Cooper pairs provides an intuitive understanding of the rogue
blueshift: analogous to the superfluid gap, there is a finite amount of energy
required to form (break) a correlated pair, leading to an increase in the amount of
energy required to excite the system, in the present case, to destroy (create) a
molecule. With this idea in mind, the rogue blueshift could serve as a
signature for the presence of Cooper pairing in a Fermi degenerate gas\cite{MAC04}.

The author acknowledges Jyrki Piilo and Kalle-Antti Suominen for helpful
discussions, as well the Acadamy of Finland for support (project 50314).


\begin{thebibliography}{99}
\bibitem{NO_ZERO}
  The exclusion of the $\k=0$ modes from the summation is
inconsequential\cite{JAV02}.
\bibitem{DK_DI}
The definition of the photoassociation rate coefficient, $\tilde{K}=dK//dI$, is
based on the standard two-body rate equation $\dot{n}=-Kn^2$.
\end{thebibliography}
\end{document}